\RequirePackage{fix-cm}
\documentclass[smallextended]{svjour3}       % onecolumn (second format)
\smartqed  % flush right qed marks, e.g. at end of proof
\usepackage{graphicx}
\usepackage{amssymb}
\begin{document}

\title{Mario Bunge on gravitational waves and the reality of spacetime%\thanks{Grants or other notes
}

\titlerunning{Bunge on gravitational waves}        % if too long for running head

\author{Gustavo E. Romero}

\institute{ Gustavo E. Romero \at
Instituto Argentino de Radioastronom{\'{i}}a (IAR, CCT La Plata, CONICET-CIC) \at
              C.C. No. 5, 1894, Villa Elisa, Buenos Aires, Argentina. \\
              Tel.: +54-221-482-4903\\
              Fax: +54-221-425-4909\\
              \email{romero@iar-conicet.gov.ar}}

\date{Received: date / Accepted: date}

\maketitle

\begin{abstract}

I discuss the recent claims made by Mario Bunge on the philosophical implications of the discovery of gravitational waves. I think that Bunge is right when he points out that the detection implies the materiality of spacetime, but I reject his identification of spacetime with the gravitational field. I show that Bunge's analysis of the spacetime inside a hollow sphere is defective, but this in no way affects his main claim.

% \PACS{ 01.70.+w  \and 04.20.Gz }

\end{abstract}
\vspace{0.3cm}

%\begin{quotation}
%
%$\varphi\'\upsilon\sigma\iota\zeta$ $\kappa\rho\'\upsilon\pi\tau\epsilon\sigma\theta\alpha\iota$ $\varphi\iota\lambda\epsilon\~\iota$.\\
%There is no dynamics within space-time itself: nothing ever
%moves therein; nothing happens; nothing changes [. . . ] one does
%not think of particles as ‘moving through’ space-time, or as ‘following
%along’ their world-lines. Rather, particles are just ‘in’
%space-time, once and for all, and the world-line represents, all at
%once, the complete life history of the particle...\\
%\begin{flushright}
%{\sl R. Geroch\footnote{R. Geroch, General relativity from A to B (University of Chicago, Chicago
%1978) pp. 20-21}.}
%\end{flushright}
%\end{quotation}  

\section{Bunge on the detection of gravitational waves}
\label{sec:1}

Mario Bunge's paper ``Gravitational waves and spacetime'' is important for at least two reasons: 1. It is the first paper to discuss some of the deep philosophical problems raised by the recent detection of gravitational waves by LIGO collaboration (Abbott et al. 2016), and 2. In his paper Bunge manifests a change in his ontological views about gravitation and spacetime. A former relationist {\it \'a la} Leibniz, Bunge now claims the identity of the gravitational field and spacetime in the light of the recent detection of gravitational waves. This amounts to some sort of spacetime realism or ``substantivalism''. I think that Bunge's analysis should be praised as timely and he deserves recognition for his brave intelectual honesty in front of the facts. His analysis and conclusions, however, are not free of some problems. The purpose of this short commentary is to discuss these problems.

On September 14th 2015, LIGO interferometric detectors were activated by a gravitational wave produced by the final inspiral of two black holes. The merger of these two objects occurred at a distance of $\sim400$ Mpc. The gravitational signal was traveling through the intergalactic space during $\sim$ 1200 Myr. Once the wave arrived to the earth it produced physical changes in the detectors of two independent instruments at Hanford, WA, and Livingston, LA (USA). Bunge's argument can be summarised like this:\\

\noindent P1. Gravitational waves activated detectors.\\
P2. Detectors react only to specific material\footnote{For Bunge, an entity is \textit{material} if it can change. Material objects, contrary to mere concepts, are changeable and can trigger changes in other objects.} stimuli.\\
P3. LIGO has detected gravitational waves.\\
 
Hence, gravitational waves are material. \\
  
\noindent P'1. Gravitational waves are ripples in spacetime.\\
P'2. Gravitational waves are material (first argument).\\

Hence, spacetime is material.\\

I think these arguments are sound. In order to argue for P'1 Bunge offers an analysis of the semantics of Einstein's equations:

\begin{equation}
R_{ab}-\frac{1}{2}R g_{ab}= \kappa T_{ab}.
\end{equation}

This is a set of ten non-linear differential equations for the metric coefficients $g_{ab}$. $R_{ab}$ is the Ricci tensor formed with second order derivatives of $g_{ab}$ and $R$ is the Ricci scalar formed by contraction of the latter tensor. $T_{ab}$ is a second order tensor that represents the properties of all  non-gravitational material fields. Finally, $\kappa$ is a constant ($8 \pi G/c^4$). All these tensors are defined over a real $C^{\infty}$-differential, 4-dimensional pseudo-Riemannian manifold. This manifold along with the metric $g_{ab}$ is supposed to represent spacetime (which can be considered a basic ontological entity). Then, according this interpretation,  Einstein's equations establish a relation between some properties of spacetime (its curvature) and the properties of matter (energy density and momentum). Solving the equations, we get the metric of spacetime, we can calculate the connection $\Gamma^c_{ab}$ formed by first order derivatives of $g_{ab}$, and then we obtain the equations of motion for test particles. If the curvature is different from zero, trajectories will depart from straight lines. If the test particle approximation cannot be ensured, the equations should be solved numerically through iterative methods in order to take into account the non-linearities. Notice that there is no gravitational field in this interpretation.  There is just spacetime and matter. Bunge's argument shows that spacetime is as material as matter. But Bunge does not stop here.   

Bunge argues that a different, equally valid interpretation of the field equations is possible, in terms of a gravitational field. This interpretation is suggested by the Newtonian limit of the theory and the comparison with the Poisson equation $\nabla^2 \phi = 4 \pi G \rho$, where $\phi$ is the potential of the gravitational field, and $\rho$ is the mass density. According to Bunge, this limit implies that the coefficients of the metric can be interpreted as the potential of the gravitational field (a view already expressed in Bunge 1967). So, Einstein's equations can be read alternatively as referring to a gravitational field or to spacetime. Since reality is unique, Bunge infers the identity of spacetime and gravitational field. I disagree.  

The coincidence of both theories in the Newtonian limit does not imply a transfer of referent from the less to the more comprehensive theory. It just implies that general relativity incorporates in its domain many results also obtained by Newton's theory, to good approximation. There is a semantical shift when we go from one theory to the other (see Bunge 1974a, b). The reference class changes, although some aspects of the formalism are recovered in the limit. In general relativity, what we call ``gravitational effects'' are due to spacetime when its curvature is different from zero. 

Although Einstein originally was inspired by Maxwell's and Lorentz's concepts of field, the final theory that resulted from his endeavours was not completely akin to Maxwell's. Einstein himself realised this after his famous debate with Willem de Sitter about dynamical empty universes (see Smeenk 2014). Spacetime has a unique ontological status in general relativity: it is an entity, which can exist by itself and, as LIGO detectors have shown, act upon matter. But spacetime can also exist in the absence of any other material entity.  Einstein recognised the ontological status of spacetime in his address delivered on May 5th, 1920 in the University of Leyden (Einstein 1920):

\begin{quotation}
 Recapitulating, we may say that according to general relativity space is endowed with physical qualities. 
\end{quotation}

The gravitational field is alien to general relativity in a similar way as classical concepts such as intrinsic angular momentum are alien to quantum mechanics. The theory, of course, can account for the phenomena we dub ``gravitational'' through curvature of spacetime. Bunge's proposal of the identity of gravitational field and spacetime leads him to confusion and error in the analysis of the interior of a hollow sphere in general relativity.

\section{The hollow sphere in general relativity}
\label{sec:2}

Bunge asks in his paper: ``What becomes of spacetime when matter vanishes, as in the case of a hollow sphere?" He argues that, as it is well known, the gravitational field in the interior of a thin shell is null (in the absence of external field). From this and his proposed identity between spacetime and gravitational field he concludes that spacetime must disappear as well from the interior of the shell. Notice that he reasons from analogy with the Newtonian case, where spacetime and gravitational field are different entities. In the Newtonian example there are no gravitational forces inside the sphere but space and time are not abolished. Actually, they are necessary to formulate the statement ``the gravitational field is zero at the coordinates such and such inside the sphere". It seems that for Bunge, in the relativistic case the absence of gravitational effects must be identified with {\it both} the absence of field {\it and} spacetime.  Actually, the field is absent from the theory from the very beginning, and spacetime still exists inside the sphere. What vanishes is the curvature of spacetime that accounts for what we call ``gravitational effects''. Let us see. 

The hollow sphere is spherically symmetric and static. By Birkoff theorem, the only solution of Einstein's equations with these characteristics has the form:

\begin{equation}
ds^2=\left(1-\frac{R_0}{r}\right) c^2dt^2 -\frac{dr^2}{1-R_0/r}-r^2d\Omega, \label{solut}
\end{equation}

\noindent where $R_0$ is a constant. Outside the shell of mass $M$, this solution reduces to Schwarzschild's:

\begin{equation}
ds^2=\left(1-\frac{2GM}{c^2r}\right) c^2dt^2 -\frac{dr^2}{1-2GM/c^2r}-r^2d\Omega, \;\;\; {\rm for}\;\;   r>R_0. \label{Schw}
\end{equation}

The interior spacetime has a metric that can be obtained from equation (\ref{Schw}) making $M=0$, since all the mass is outside the region under consideration. Then, 

\begin{equation}
ds^2= c^2dt^2 -dr^2-r^2d\Omega, \;\;\; {\rm for}\;\;   r<R_0.
\end{equation}
This is Minkowski metric. This means that spacetime exists inside the shell, but its metric is flat and hence there are no gravitational effects, exactly as in the Newtonian case. Spacetime does not disappear, just curvature vanishes, and then test particles cannot experience any deviation that might be attributed to gravitation. 

\section{The reality of spacetime}
\label{sec:3}

Why has Bunge missed this point after correctly recognising the physical reality of spacetime? I think that he is not still free from his longly espoused and recently abandoned relationism (Bunge 1977). Bunge seems to think that in the absence of matter, and consequently in the absence of relations among material bodies, spacetime cannot survive. A staunch relationist about spacetime would say exactly the same thing. Bunge is not taking seriously enough his own conclusion enunciated above: spacetime is material. And as a material entity, spacetime can exist in absence of matter -- just as gravitational waves show us it is the case\footnote{Spacetime is a 4-dimensional entity. This means that in 4-dimensions spacetime does not change. What we call changes are asymmetries in 3D slices of a 4 dimensional object (see Romero 2013). In order to achieve full consistency, Bunge should abandon the last residuum of relationism in his ontology and seriously consider whether ontological realism about spacetime is compatible with another cherished metaphysical doctrine: presentism. I have argued extensively against presentism elsewhere (Romero 2015, 2017).  }. 

To embrace the reality of spacetime is to accept that it is a material entity.  This materiality is responsible for the non-linear nature of Einstein's theory. All kind of material entities can interact with spacetime through curvature, and this includes spacetime itself. This is a lesson that Einstein understood from de Sitter when they discussed the cosmological implications of the theory and a lesson that Bunge should assimilate if he wants to include spacetime in the right place within his vast ontology.

%\section{Against presentism}
%\label{sec:4}

\section{Final remarks}

General relativity is a theory about the interactions of spacetime and other material systems. The theory is eliminative with respect to the old concept of gravitational field. Spacetime curvature is responsible for the deviation of test particles from straight trajectories and replaces the old idea of a gravitational field defined on space and time acting locally. Spacetime itself, as Bunge points out, is a material entity. This opens the door to the important problems of its nature and composition and suggests that approaches based on field theory  might be seriously flawed. The nineteenth and twentieth centuries have seen the rise of field theories in physics. But in order to understand the inner nature of spacetime, we should push, perhaps, even beyond.   

%\section*{Appendix: Axiomatic quantum mechanics}
%\label{Appendix}

\begin{acknowledgements}
I thank Mario Bunge for stimulating discussions. My research on gravitation is supported by grant PIP 0338 (CONICET) and grant
AYA2016-76012-C3-1-P (Ministro de Educaci\'on, Cultura y Deporte,
Espa\~na).
\end{acknowledgements}




\newpage

\section*{Gustavo E. Romero} Full Professor of Relativistic Astrophysics at the University of La Plata and Superior  Researcher of the National Research Council of Argentina. A former President of the Argentine Astronomical Society, he has published more than 350 papers on astrophysics, gravitation, and the foundations of physics. Dr. Romero has authored or edited 10 books (including {\sl Introduction to Black Hole Astrophysics}, with G.S. Vila, Springer, 2014). His main current interest is on black hole physics and ontological problems of spacetime theories.

\end{document}